\documentclass{elsart}
\usepackage{amssymb,psfig,subfigure,graphicx}
\usepackage{amsmath}
\usepackage{psfrag,epsfig}
\newcommand{\eps}{\varepsilon}
\newcommand{\ket}[1]{\left | #1 \right\rangle}
\newcommand{\comm}[2]{\left[ #1, #2 \right]}
\newcommand{\myfrac}[2]{\genfrac{}{}{0pt}{}{#1}{#2}}
\newcommand{\la}{\lambda}

\def\dwn{\downarrow}
\def\up{\uparrow}
\def\d{\dagger}
\def\beq{\begin{equation}}
\def\eeq{\end{equation}}
\def\bal{\begin{align}}
\def\eal{\end{align}}
\def\dwn{\downarrow}
\def\up{\uparrow}
\def\d{\dagger}

\def\CC{{\rm\kern.24em \vrule width.04em height1.46ex depth-.07ex
\kern-.30em C}}
\def\P{{\rm I\kern-.25em P}}
\def\RR{{\rm
         \vrule width.04em height1.58ex depth-.0ex
         \kern-.04em R}}
\def\id{{\rm 1\kern-.22em l}}
\def\ZZ{{\sf Z\kern-.44em Z}}
\def\NN{{\rm I\kern-.20em N}}
\def\n{\hat{n}}

\def\K{\hat{K}}
\def\S{\hat{S}}

\begin{document}
\begin{frontmatter}
\title{Quasi-classical descendants of disordered 
vertex models with boundaries }
\author[dmfci]{Antonio Di Lorenzo}, 
\author[dmfci]{Luigi Amico},
\author[tokyo]{Kazuhiro Hikami}, 
\author[dmfci]{Andreas Osterloh}, and 
\author[dmfci]{Gaetano Giaquinta}
\address[dmfci]{NEST-INFM $\&$Dipartimento di 
Metodologie Fisiche e Chimiche (DMFCI), 
        Universit\`a di Catania, viale A. Doria 6, I-95125 Catania, Italy} 
\address[tokyo]{ Department of Physics, Graduate School of Science, University 
of Tokyo, Hongo 7-3-1, Bunkyo, Tokyo 113-0033, Japan}
\begin{abstract}
We study  descendants of 
inhomogeneous vertex models with boundary reflections  when the 
spin-spin scattering 
is assumed to be quasi--classical. This corresponds to consider certain power 
expansion of the boundary-Yang-Baxter equation  (or reflection equation). 
As final product, integrable $su(2)$-spin  chains interacting 
with a long range with $XXZ$ anisotropy are obtained. The spin-spin 
couplings are non uniform, and a  non uniform tunable 
external magnetic field is applied;
the latter can be obtained  
when the boundary conditions  are assumed to 
be quasi-classical as well. The exact spectrum is achieved by 
algebraic Bethe ansatz.
Having realized the $su(2)$ operators in terms of fermions, 
the class of models we found  turns out to describe
confined fermions with pairing force  interactions. The class of models 
presented in this paper 
is a one-parameter extension of certain Hamiltonians constructed 
previously. Extensions to $su(n)$-spin open chains  are discussed. 
\end{abstract}
{\it PACS} N.  02.30.Ik, 75.10.Jm
\end{frontmatter}
\maketitle

\section{Introduction}
Integrable vertex models (VM) in two dimensional 
classical statistical mechanics 
are the common seed % proved to be a fertile ground for 
of many relevant exactly-solved quantum models 
in one dimension~\cite{Baxter82,Gaudin83}. 
Famous examples are  the $XXX$, $XXZ$, and $XYZ$ Heisenberg 
chains that find more and more applications in contemporary physics.
The key toward this  powerful synthesis 
is to notice that the ``scattering'' of 
the degrees of freedom of both the VM and the spin chains  is described  
by the same matrix.  The Quantum Inverse Scattering 
Method (QISM) exploits  this  fact systematically~\cite{Korepin93}.
The method relies on the observation that transfer matrices $\hat{t}=Tr (T)$  
span a one-parameter family of commuting operators  
if  a (scattering) matrix $R$ exists such that 
$T$, $R$ satisfy the celebrated 
Quantum Yang-Baxter equation. The equivalence 
between VM and Heisenberg chains consists 
in the fact that these models have the same $R$--matrix. 
Due to the property of the scattering 
$R(u,v)=R(u-v)\;,\quad \forall u, v \in \CC$ 
the integrability of VM is 
preserved if disorder is added at each lattice site such that the scattering 
``wave momenta'' $u$, $v$ result to be shifted arbitrarily. 
In this case, however, it is difficult to extract a Hamiltonian. 
A route to simplify the problem is to resort the 
so called ``quasi-classical'' limit 
of the QISM. The term ``quasi-classical'' here indicates 
that the scattering between the degrees of freedom of the model is assumed to 
be quasi-classical. 
Quantitatively, this means that a parameter $\eta$ 
does exist such that   
the scattering matrix is of the form 
$R(u)\propto\id \otimes \id + \eta r(u)$ in the 
limit $\eta\rightarrow 0$ ($\eta$ plays the role of $\hbar$).
The quantity $r(u)$  
fulfills the classical Yang-Baxter equation (that is a restatement 
of the Jacobi identity for the Poisson brackets of suitable 
action-angle variables). It is worthwhile to mention, however,
that the systems obtained by this quasi-classical expansion consist of  
{\it quantum} spins (by no means quasi-classical). 
The quasi-classical expansion of the transfer 
matrix of disordered  VM (in the lowest spin 
representation) is non trivial and 
it produces the Gaudin's magnet 
Hamiltonians\cite{Gaudin76,Hikami92} 
containing a long range spin  interaction (in contrast with the 
range of the Heisenberg chains which involves nearest neighbour spins). \\
A richer  variety of integrable models by QISM comes from  imposing 
non trivial  boundary conditions different from the periodic ones. 
Twisted  boundary conditions, for example, 
imposed to the six vertex model~\cite{Sklyanin89} produce 
the Gaudin magnet  in 
a  non-uniform local magnetic field, which is very  
important for physical applications. 
In fact having realized  the (pseudo)spin algebra in terms of fermions  
the $XXX$ Gaudin Hamiltonians in a non 
uniform magnetic field are 
the constants of the motion of the BCS model\cite{Cambiaggio97} 
that describes pairs of electrons 
(in time reversed states) interacting with  a long range uniform 
pairing coupling. The  exact solution of the BCS model was found 
long ago by Richardson\cite{Richardson63a} 
and rediscovered recently.  In particular it was used to study small metallic 
grains~\cite{RalphvonDelft01,Mastellone98}; the picture was merged in the 
scenario of QISM in the Ref.~\cite{Amico01}.
Connections with WZNW models in field theory 
have been deeply investigated~\cite{Sierra00} 
based on the relation  between solution of KZ equation and
Gaudin model found in Refs.~\cite{Babujian93,Reshetikhin95}. 
The class of pairing Hamiltonians was generalized by investigating 
the quasi-classical expansion of the disordered 
twisted six vertex model  with $XXZ$ $R$-matrix. 
In terms of fermions this class of Hamiltonians 
represents interacting electron pairs with 
certain non-uniform long-range 
coupling strengths~\cite{Amico01a,Amico01b,Dukelsky01,vonDelft01}. 
Twisted   rings can be cut to open chains and loops
include two reflections at the boundaries. The possibility 
to include such reflections in 
integrable theory was founded and systematically 
investigated by Sklyanin~\cite{Sklyanin88}.  
The quasi-classical  limit of the disordered 
six vertex model with boundaries was investigated 
first by one of the authors~\cite{Hikami95,Hikami95a}. 
This led to a model where the spin couplings  
contain an additional parameter with respect 
to the original Gaudin magnet, and in a {\it vanishing
 external} magnetic field (see Eq.~(\ref{tau-classical}) and the relative  
discussion  below of it). 
In the present work, we proceed along this line.
We still consider an inhomogeneous six-vertex model 
with boundary reflection, following closely Ref.~\cite{Hikami95}. 
By properly choosing the reflection 
parameters, we introduce an external   
non-uniform magnetic field of {\it tunable} strength in the
Hamiltonian.
The trick consists in the assumption that also the boundary conditions 
have a quasi-classical expansion (see Eq.~(\ref{boundary-expansion}) and 
Sec.~\ref{nonclassical}). At best of our knowledge, 
this idea is pursued 
for  the first time in the present paper.
In the following we summarize the main results obtained in the bulk of the 
paper. 
The class of spin-$S_j$ ($j=1,\dotsc,N$) models that we find  
 has Hamiltonian of the form 
\begin{equation}
H= \sum_j 2 h_j\S_j^z - 
\sum_{\myfrac{j,k}{j\neq k}}
\left [J_{jk}^{(z)} \S_j^z \S_k^z 
+ \left ( J_{jk}\S_{j}^+ \S_{k}^- + h.c.\right) \right ]
\;.
\label{Themodel}
\end{equation}
We agree that the latin indices $j, k$ will run from 1 to $N$, where 
$N$ is the number of spins.  
The operators $S^\pm\, , \; S^z$ are $su(2)$ operators.
The couplings are
\begin{align}
\label{couplings}
J_{jk}^{(z)}&= I_{jk}~ 
{\bigl(\cosh(2pz_j)+\cosh(2pz_k)-2\cos{(2pt)}\bigr)}\nonumber \\
J_{jk}&= I_{jk}~\bigl({\sinh[p(z_j-it)]\sinh[p(z_k+it)]}
\bigr)  \nonumber \\
I_{jk}&= J(\S^z)\frac{h_j-h_k}{\cosh(2pz_j)-\cosh(2pz_k)}\;, \nonumber \\
J(\S^z)&= J \left(1-J\S^z\right)^{-1}
\end{align}
where $\S^z$ is the total $z$-component of the  spin.
The quantities  $h_j, z_j$ are two arbitrary sets of real parameters; 
$t$ is also a real arbitrary parameter and it directly comes 
from the boundary 
terms (see Eq.~(\ref{boundarymatrix}) with $\xi=it$);
finally, $p$ can be $1,i$ or can be tending  to zero corresponding to
hyperbolic, trigonometric, and  rational couplings, respectively.\\
The eigenstates  
in the sector with total $z$-component of the spin  $S^z=\sum_j S_j - M$, are  
\begin{equation}
\ket{\Psi}=\prod_{\alpha=1}^M \S^-(e_\alpha)  \ket{H}
 \;,
\label{eigenvec}
\end{equation}
where 
\begin{equation}
\S^-(u)= \sum_j
\frac{\cosh[p(u+z_j+2 i t)]-\cosh[p(u-z_j)]}
{\cosh(2pu)-\cosh(2pz_j)}
\S^-_j 
\end{equation}
and
\begin{equation*}
\ket{H} = \bigotimes_{j=1}^N \ket{S_j^z = S_j}\;.
\end{equation*}
The corresponding eigenvalues are 
\begin{equation}
\label{eigenval}
{\mathcal E}=\sum_j 2 h_j \tau_i
\end{equation}
where
\begin{equation}
\tau_j= 
S_j \left(1  - J(S^z)\sum_{k\neq j} S_k \frac{1 - x_j x_k}{x_j - x_k}
+ J(S^z)\sum_\alpha 
\frac{1 - x_j \la_\alpha}{x_j - \la_\alpha}
\right) \;, 
\end{equation}
where we defined
\begin{equation}\label{reparameter}
\frac{1+x_j}{1-x_j} = \cosh(2pz_j) - \cos(2pt)\;\;,\;\;
\frac{1+\la_\alpha}{1-\la_\alpha} = \cosh(2pe_\alpha) - \cos(2pt) 
\end{equation}
The \emph{rapidities} $\la_\alpha$ satisfy, in the sector having 
total $z$-component of the spin  $S^z$, the Bethe equations: 
\begin{equation}
\begin{split}
&\sum_{\beta\neq\alpha} \frac{1}{\la_\alpha-\la_\beta} 
- \sum_j \frac{S_j}{\la_\alpha - x_j} + \\
&+\frac{1}{2J(S^z)}\left(\frac{1+J(S^z)(1+S^z)}{1+\la_\alpha} 
+\frac{1-J(S^z)(1+S^z)}{1-\la_\alpha}\right) =0\;.
\end{split}
\end{equation}
We point out that the dependence on the reflection 
parameters comes only in the  couplings  
and eigenvectors (Eqs.~(\ref{couplings})(\ref{eigenvec})), 
while the eigenvalues depend on $t$ only implicitely 
(through Eq.\eqref{reparameter}). 
\\
The rational limit of the models is recovered for $p\to 0$.
\\
For $t=0$ and $pt=\pi/2$, the models 
reduce to the ones that we presented in Ref.~\cite{Amico01a} 
(see section \ref{noreflection}).  
Thus the  class of models we discuss in the present paper is  
a one-parameter extension of the former class.   

Using the fermionic realizations of $su(2)$ the Hamiltonian~(\ref{Themodel}) 
can be rephrased to describe confined fermions interacting  
with  pairing and exchange forces (see Eq.~(\ref{ham})).

The paper is organized as follows. 
In the next section we summarize the main ingredients  
of the inverse scattering of 
 VM with boundaries. In section III 
we construct the integrable models 
we deal with together with their exact solution. 
In section IV we use the 
fermionic realization of the $su(2)$ 
algebra to rewrite the Hamiltonians in a 
second quantized form. Section V is devoted to final remarks. 
In appendix~\ref{VM} we review basic properties of VM. 
In appendix~\ref{genrefl}
we prove the integrability of a class of models 
when a more general (off-diagonal) reflection at the boundary 
is applied (see Eqs.~(\ref{off-diagonal-K})~--~(\ref{off-diagonal-tau})).
We also discuss a generalization to $su(n)$ case in
appendix~\ref{sec:sun}.

%%%%%%%%%%%%%%%%%%%%%%%%%%%%%%%%%%%%%%%%%%%%%%%%%%%%%%%%%%%%%%%%%%%%%%%%%%%%%

\section{Integrable boundary conditions}\label{boundary-conditions}
In this section, we review how the QISM is applied to 
VM, in order to obtain a family of commuting transfer matrices. 
VM describe a system of interacting classical objects 
on a two dimensional lattice. As described in 
appendix \ref{VM}, the partition function of the system can be written as 
\(Z=\text{Tr}\{\hat{t}(1) \dotsm \hat{t}(K)\}\), where 
$\hat{t}(i)$ are operators in some appropriate \emph{many-body} 
linear space (in the sense that it 
is the direct product of $N$ elementary linear spaces). 
The VM is exactly solvable if $\comm{\hat{t}(i)}{\hat{t}(i')}=0$. 
Usually, it is assumed that the dependence on the $i$-th row of the lattice 
comes through a parameter $u_i$, which takes values on some  
domain of the complex plane. 
Then the requirement for exact solvability becomes 
$\comm{\hat{t}(u)}{\hat{t}(v)}=0$, $\forall u,\,v$ belonging 
to the domain. 

The QISM provides a way of constructing classes of commuting 
operators $\hat{t}(u)$, 
finding their eigenvalues and their common eigenstates, and 
extracting Hamiltonians whose integrals of motions are $\hat{t}(u)$.  
The QISM is a procedure which starts from the $R$-matrix  
and from the Lax operator 
to yield the transfer matrix 
$\hat{t}(u)$. 
From the transfer matrix, a class of 
Hamiltonians can be extracted in various ways, to be depicted below. 
The QISM has a built-in Algebraic Bethe Ansatz (ABA) which provides the 
diagonalization of the $\hat{t}(u)$, and hence of the Hamiltonian. 

The $XXZ$ $R$-matrix is 
\begin{equation}
R(u,v)=
\begin{pmatrix}
a(u,v)& 0& 0&0 \\
0 & b(u,v)&c(u,v)&0 \\
0 & c(u,v)& b(u,v)&0 \\
0 & 0& 0&a(u,v)  \end{pmatrix}\;,
\end{equation} 
where 
\begin{equation*}
a(u,v)=\sinh{[p(u\!-\!v\!+\!\eta)]}/p \;, \,
b(u,v)=\sinh{[p(u\!-\!v)]}/p \;, \,
c(u,v)=\sinh{\!(p\eta)}/p\,.
\end{equation*}
It is connected to the $\check{R}$-matrix defined for VM by 
\(R(u,v)= {\mathcal P}_{12} \check{R}(1,2)\), where 
\[{\mathcal P}_{12}=\begin{pmatrix}1&0&0&0\\0&0&1&0\\
0&1&0&0\\0&0&0&1\end{pmatrix}\] 
is the permutation operator, and 
it is assumed 
that the dependence of $R$ upon the rows comes through a parameter assigned to 
each row. \\
The corresponding Lax operators are
\begin{equation}
L_j(u)= \frac{1}{p}
\begin{pmatrix}
\sinh[p(u + \eta \S^z_j)] & \sinh(p\eta)~\S^-_j\\
\sinh(p\eta)~\S^+_j & \sinh[p(u - \eta \S^z_j)]
\end{pmatrix}\;.
\end{equation}
Here $p$ is the anisotropy parameter, in the sense that, 
when $p\neq 0$ --- in which case one can put either $p=1$ or $p=i$ --- 
the QISM yields a Hamiltonian with $XXZ$-type couplings, 
while in the limit $p\to 0$, the hyperbolic/trigonometric functions 
reduce to rational ones, and the QISM generates 
a Hamiltonian having $XXX$ couplings; $\eta$, instead, 
is the so-called quantum parameter which plays the role of $\hbar$;
as we shall later see, 
it gives the degree of deformation of the 
\emph{classical} algebra $su(2)$ into the \emph{quantum} algebra $su_q(2)$. 
We remark that the terminology is somehow misleading:  
since we associate the algebra $su(2)$ with spins, 
realized either by true spins or 
by pairs of time-reversed electrons, 
in the limit $\eta\to 0$ we obtain genuine \emph{quantum} 
Hamiltonians.

The Lax operators act on the \emph{auxiliary} 
two-dimensional vector space ${\mathcal V}$, and on 
the \emph{quantum} space ${\mathcal H}_n$. 
They obey the fundamental Yang-Baxter relation Eq.\eqref{fund}, which 
in terms of the $R$-matrix now reads
\begin{equation}
\stackrel{ }{R(u-v)} {\stackrel{1}{L}}_j(u-z_j)  \stackrel{2}{L}_j(v-z_j)
= \stackrel{2}{L}_j(v-z_j) \stackrel{1}{L}_j(u-z_j) \stackrel{ }{R (u-v)}\;,
\label{fundamental}
\end{equation}
Due to the additive property of the $R$-matrix $R(u,v)=R(u-v)$,
parameters $z_j$ taking into account  {\it on-site} disorder through 
the lattice can be introduced. \\
As customary \({\stackrel{1}{L}}_j(u) = L_j(u)\otimes \id\), and 
\({\stackrel{2}{L}}_j(u) = \id\,\otimes L_j(u)\); 
the external product is meant between two copies of the space ${\mathcal V}$, while the 
multiplication of the elements of $L_j$, which are operators on 
${\mathcal H}_j$, is an internal product. 
The relation \eqref{fundamental} is actually obeyed only for 
$1/2$ spins, i.~e.~ for $dim(\mathcal{H}_j)=2$. The order of the 
representation (that is the dimension of $\mathcal{H}_j$) 
can be extended to larger values, keeping the dimension of ${\mathcal V}$ fixed to 2; 
however, one has to renounce to the algebra $su(2)$, and introduce rather 
the \emph{quantum} algebra $su_q(2)$, which ensures that the relation 
\eqref{fundamental} is obeyed whatever is the representation of the algebra. 
The parameter $q$ is related to the 
parameters $p$ and $\eta$ by $q=exp(p\eta)$. The commutation rules are
\begin{equation}
\comm{\S_j^z}{\S_j^\pm} = {\S_j^\pm} \;\;;\;\;
\comm{\S_j^+}{\S_j^-} = \frac{\sinh(2p\eta\S_j^z)}{\sinh(p\eta)}\;.
\end{equation}
In the quasi-classical limit $\eta\rightarrow 0$, 
or in the isotropic limit $p\to 0$, 
$su_q(2)$ reduces to $su(2)$.
\\
Next, we consider 
the monodromy matrix \(T(u)\equiv L_1(u-z_1) \dotsm L_N (u-z_N)\).  
We have an internal product over ${\mathcal V}$ and an external one 
over \({\mathcal H}_j\) and \(\mathcal{H}_{j'}\); thus $T(u)$
is an operator over 
\({\mathcal V}\otimes {\mathcal H}_1\otimes\dotsm \otimes \mathcal{H}_N\).
It has the form 
\[T(u)=
\begin{pmatrix}
A(u)&B(u)\\
C(u)&D(u)
\end{pmatrix}\;,\]
with $A,~B,~C,~D$ operators over $\mathcal H = \bigotimes_j {\mathcal H}_j$. 

The local relation~(\ref{fundamental}), and the ultra-locality 
property, \(\comm{L_j^{ab}(u)}{L_k^{cd}(v)} = 0\) for $j\neq k$, imply that 
$T(u)$ fulfills  
the global Yang-Baxter equation  
\begin{equation}
\stackrel{ }{R(u-v)} {\stackrel{1}{T}}(u)  \stackrel{2}{T}(v)
= \stackrel{2}{T}(v) \stackrel{1}{T}(u) \stackrel{ }{R (u-v)}\;,
\label{non-local}
\end{equation}  
In the case of periodic boundary conditions (on the auxiliary matrix space), 
the quantities $Tr\{ \,T(u)\} = A(u)+D(u)$, 
where the trace is on the auxiliary space ${\mathcal V}$,  
generate a one parameter commuting family 
of operators which underlies an integrable model. 
\\
Remarkably, the property of integrability is preserved for a wider class 
of non-trivial boundary conditions. Integrable boundary conditions are 
introduced by 
the so-called ``boundary $K$-matrices'' satisfying the reflection 
equations~\cite{Cherednik84,Sklyanin88}
\begin{align}
\nonumber
R(u-v) &\stackrel{1}{K}_-(u)~ R(u+v) \stackrel{2}{K}_-(v)=
\\[.2cm] & \label{reflminus}
 \stackrel{2}{K}_-(v) R(u+v) \stackrel{1}{K}_-(u) R(u-v) 
\end{align}
\begin{align}
\nonumber
R(-u+v) &{\stackrel{1}{K}_+^t}(u) R(-u-v-2 \eta) 
{\stackrel{2}{K}_+^t} (v)= 
\\[.2cm] &\label{reflplus}
{\stackrel{2}{K}^t_+ (v)} R(-u-v-2 \eta ) 
{\stackrel{1}{K}^t_+} (u) R(-u+v)  
\end{align}
Among the solutions of the reflection equations, 
we consider the diagonal ones, 
which yield Hamiltonians preserving the total $z$-component of the spin  $\S^z$. 
They depend upon the free parameters $\xi_\pm$
\begin{align*}
K_-(u)&= K(u,\xi_-)\\
K_+(u)&= K(u+\eta,\xi_+) 
\end{align*}
where
\begin{equation}
K(u,\xi)= \frac{1}{p}
\begin{pmatrix}
\sinh{[p(u+\xi)]} &0 \\
0& -\sinh{[p(u-\xi)]}
\end{pmatrix}
\label{boundarymatrix}
\end{equation}
The family of commuting transfer matrices is 
\begin{equation}
\hat{t}(u) = Tr\{ K_+(u) U(u)\}
\end{equation}
where \(U(u)=~T(u) K_- (u) T^{-1}(-u) = \begin{pmatrix}
{\mathcal A}(u)&{\mathcal B}(u)\\ {\mathcal C}(u)&{\mathcal D}(u) 
\end{pmatrix}\). 
The \emph{inverse} of the monodromy matrix is defined as \cite{Korepin93}
\[T^{-1}(-u)={\boldsymbol\sigma^y} T^t(-u+\eta) 
{\boldsymbol\sigma^y}~\text{det}_q^{-1} T(-u+\eta/2)\;,\]
where ${\boldsymbol\sigma^y}$ is the Pauli matrix in the representation where 
${\boldsymbol\sigma^z}$ is diagonal, and 
$\text{det}_q T(u) = A(u-\eta/2) D(u+\eta/2)-C(u-\eta/2) B(u+\eta/2)$ 
is the quantum determinant, which is a $su(2)$-number. 

The eigenvectors of $\hat{t}(u)$, 
in the sector with $S^z = \sum_j S_j - M$, 
are given by~\cite{Sklyanin88} 
\[\prod_{\alpha=1}^M {\mathcal B}(e_\alpha) \ket{H}\,,\] 
where \[\ket{H}=\bigotimes_j \ket{S_j^z=S_j}\] 
is the pseudo-vacuum state having all maximum $S_j^z$ eigenvalues. 
The eigenvalues are 
\begin{equation}
\begin{split}
&t(u) = \frac{\sinh[2p(u+\eta)]}{\sinh[p(2u+\eta)]}
\bigl(\cosh[2p(u+\sigma)]-\cosh(2p\delta)\bigr) \times \\
&\times 
\left[\prod_{\alpha=1}^M \frac{\sinh[p(u-e_\alpha-\eta)]\sinh[p(u+e_\alpha)]}
{\sinh[p(u+e_\alpha+\eta)]\sinh[p(u-e_\alpha)]} \right] 
a(u) d(-u+\eta)
+\\
&-\frac{\sinh(2pu)}{\sinh[p(2u+\eta)]}
\bigl(\cosh[2p(u-\sigma+\eta)]-\cosh(2p\delta)\bigr)\times\\
&\times
\left[\prod_{\alpha=1}^M 
\frac{\sinh[p(u-e_\alpha+\eta)]\sinh[p(u+e_\alpha+2\eta)]}
{\sinh[p(u+e_\alpha+\eta)]\sinh[p(u-e_\alpha)]} \right] 
a(-u+\eta) d(u)
\;,
\end{split}
\end{equation}
where we put $\xi_+ + \xi_- = 2\sigma$, $\xi_+ - \xi_- = 2\delta$ and  
\(a(u) = \prod_{j=1}^N \sinh[p(u-z_j+\eta S_j)]/p\), 
\(d(u) = \prod_{j=1}^N \sinh[p(u-z_j-\eta S_j)]/p\).
The $e_\alpha$ satisfy the Bethe equations: 
\begin{align}\nonumber
&\frac{\cosh[2p(e_\alpha+\sigma)] -\cosh(2p\delta)}
{\cosh[2p(e_\alpha-\sigma+\eta)]-\cosh(2p\delta)}
\prod_{\beta \neq \alpha}\frac{
\sinh[p(e_\alpha-e_\beta - \eta)] \sinh[p(e_\alpha+e_\beta)]}
{\sinh[p(e_\alpha-e_\beta + \eta)]\sinh[p(e_\alpha+e_\beta+2\eta)]}
= \\[.3cm]
&\prod_{j}{\frac{
\sinh[p(e_\alpha-z_j-\eta S_j)]~\sinh[p(e_\alpha+z_j-\eta(S_j+1))]}
{\sinh[p(e_\alpha-z_j+\eta S_j)]~\sinh[p(e_\alpha+z_j + \eta(S_j-1))]}}
\label{genbethe}
\end{align}

The final step to obtain integrable models from the procedure above 
is to observe that transfer matrices can be used as generating functional 
of Hamiltonians. A possibility is 
\begin{equation}
\label{log}
\left . H\equiv \frac{\partial}{\partial u} \ln \hat{t}(u) \right |_{u=u_c}.
\end{equation}
In the homogeneous case $z_j=0 \; \forall j$, Heisenberg  Hamiltonians with 
nearest-neighbour interaction  are obtained for $u_c=0$.
The presence of disorder makes the application of Eq.\eqref{log} 
quite difficult.
A particular value of $u_c$ for which the calculations can be done 
is $u_c\to \infty$; in this case the 
interaction in the Hamiltonian is long range~\cite{deVega84}. 
Another way to face the problem is to resort to 
the quasi-classical expansion. The trick consists in  obtaining a set of 
commuting operators as coefficients of the power-$\eta$ expansion 
of $\hat{t}(u)$ --- from which a 
Hamiltonian turns out  can be built as a polynomial.
The quasi-classical expansion of the $R$--matrix 
and Lax operators gives
\begin{align*}
L_j(u) &=\frac{\sinh(pu)}{p} 
\left[\id +\eta\,l^{(1)}_j (u) +\eta^2 l^{(2)}_j(u) 
+ {\it O}(\eta^3) 
\right]\;,\\
R(u) &= \id\otimes\id + \eta\, r(u) + {\it O}(\eta^2)\;,
\end{align*}
where $\id$ is the $(2\times 2)$ identity and $r(u)$ reads
\[r(u)=\begin{pmatrix}\cosh(pu)\;&\;0\;\;&\;0\;&\;0\\0\;&\;0\;&\;1\;&\;0\\
0\;&\;1\;&\;0\;&\;0\\0\;&\;0\;&\;0\;&\;\cosh(pu)\end{pmatrix}\; .\] 

Unfortunately, if one wants to extend the results to spins higher 
than 1/2, one has to increase the dimension of the auxiliary space 
accordingly~\cite{Babujian82}. There is, though, the remarkable exception of isotropic 
models, i.e.~the ones obtained in the $p\to 0$ limit, for which the quantum 
algebra reduces to $su(2)$. The $XXX$ model with higher spin was 
introduced in Refs.\cite{Babujian82,Takhtajan82}. 
A central point of the approach is that, 
the quasi-classical limit $\eta\to 0$ 
reduces  the quantum algebra  to \emph{ordinary} $su(2)$, whatever is
the dimension of the quantum space ${\mathcal H}_j$. 
This implies that the operators thus found are realized through 
\emph{spin} operators $\S_j^z$, $\S_j^\pm$, 
and that they commute with each other for \emph{arbitrary} spins.

%%%%%%%%%%%%%%%%%%%%%%%%%%%%%%%%%%%%%%%%%%%%%%%%%%%%

\section{Quasi-classical expansion of 
vertex models with boundary reflections}\label{integrability}
In this section we investigate systematically the power $\eta$-expansion 
of the transfer matrix of disordered vertex models with boundary. As final 
product of this procedure  we obtain a class of integrable models describing 
interacting quantum spins with non uniform long range interaction.  
   
The quasi-classical expansion of the transfer matrix reads  
\begin{equation}
\hat{t}(u)= \hat{\tau}^{(0)} + 
\eta\,\hat{\tau}^{(1)}(u) + \eta^2 \,\hat{\tau}^{(2)} (u) +{\it O}( \eta^3)
\label{quasi-transfer}
\end{equation}
The property  $[\hat{t}(u),\hat{t}(v)]=0$ ensures the existence of hierarchy 
of integrable models in the quasi-classical expansion. 
We have indeed 
\[[\hat{t}(u),\hat{t}(v)]=\sum_{l=0}^{\infty} \eta^l C_l(u,v) = 0\;,\]
which implies $C_l(u,v)=0$. 
We give the first five terms
\begin{align}\label{commexp}
C_0(u,v)=&[\hat{\tau}^{(0)}(u),\hat{\tau}^{(0)}(v)]\;,\nonumber\\
C_1(u,v)=&[\hat{\tau}^{(0)}(u),\hat{\tau}^{(1)}(v)] +
[\hat{\tau}^{(1)}(u),\hat{\tau}^{(0)}(v)]\;, \nonumber \\ 
C_2(u,v)=&[\hat{\tau}^{(0)}(u),\hat{\tau}^{(2)}(v)]
+[\hat{\tau}^{(2)}(u),\hat{\tau}^{(0)}(v)]
+[\hat{\tau}^{(1)}(u),\hat{\tau}^{(1)}(v)]\;,\nonumber \\
C_3(u,v)=&[\hat{\tau}^{(0)}(u),\hat{\tau}^{(3)}(v)]
+[\hat{\tau}^{(3)}(u),\hat{\tau}^{(0)}(v)]
+[\hat{\tau}^{(1)}(u),\hat{\tau}^{(2)}(v)]+
[\hat{\tau}^{(2)}(u),\hat{\tau}^{(1)}(v)]\;,\nonumber \\
C_4(u,v)=&[\hat{\tau}^{(0)}(u),\hat{\tau}^{(4)}(v)]+
[\hat{\tau}^{(4)}(u),\hat{\tau}^{(0)}(v)]+ \nonumber \\
&[\hat{\tau}^{(1)}(u),\hat{\tau}^{(3)}(v)]+
[\hat{\tau}^{(3)}(u),\hat{\tau}^{(1)}(v)]+
[\hat{\tau}^{(2)}(u),\hat{\tau}^{(2)}(v)]\;. 
\end{align}
From the expansion above, one finds that the first non-trivial term 
$\hat{\tau}^{(n)}(u)$ (i.e.~which is not 
just a  $\CC$-number or an invariant of the algebra) gives rise to a family 
of commuting operators. 
For example, if~\(\hat{\tau}^{(0)}=\CC-number\) (as usually is the case) 
the first class of integrable models is generated 
by $[\hat{\tau}^{(1)}(u),\hat{\tau}^{(1)}(v)]=0$. 
In the presence of boundary conditions corresponding to generic choice 
of $\xi_\pm$ it turns out that $\hat{\tau}^{(1)}(u)$ is non-trivial. 
This is not what we wish, since $\hat{\tau}^{(1)}(u)$ only 
contains non-interacting spins.\footnote{In general, the $n$-th order terms 
will contain up to $n$-body terms.}
In the next section we will see how the parameters $\xi_\pm$ can be 
suitably chosen to yield ``trivial'' $\hat{\tau}^{(1)}(u)$, such that the 
first non-trivial term in the quasi-classical expansion of $\hat{t}$ 
will be $\hat{\tau}^{(2)}(u)$, which yields a spin-spin interaction. 

The $\eta$ expansion of the Lax operators is 
\begin{equation}
L_j(u) \simeq \frac{\sinh(pu)}{p}\left[\id + \eta\,l_j^{(1)}(u) + 
\eta^2\,l_j^{(2)}(u)\right]\;,
\end{equation}
where
\bal
l_j^{(1)}(u) &=  p \coth(pu) \S_j^z {\boldsymbol\sigma^z} 
+ \frac{p}{\sinh{(pu)}} 
(\S_j^+ {\boldsymbol\sigma^-} + \S_j^- {\boldsymbol\sigma^+}) 
\;,\nonumber\\
l_j^{(2)}(u) &= \frac{p^2}{2} \left(\S_j^z\right)^2\id \;.
\end{align}
Thus, the monodromy matrix is: 
\begin{align*}
&T(u) \simeq ~P(u)\biggl\{\id\!+\!\eta\sum_j \frac{p}{\sinh{(pu_j^-)}}
\left(\cosh{(pu_j^-)} 
\S_j^z {\boldsymbol\sigma^z}\!+\!\S_j^+ 
{\boldsymbol\sigma^-}\!+\!\S_j^- {\boldsymbol\sigma^+}\right)
+ \\
&+ p^2 \eta^2 \biggl[\sum_{j<k} 
\frac{\cosh{(pu_j^-)}\cosh{(pu_k^-)} \S_j^z \S_k^z \id + 
\S_j^+ \S_k^- {\boldsymbol\sigma^-} {\boldsymbol\sigma^+} 
+ \S_j^- \S_k^+  {\boldsymbol\sigma^+} {\boldsymbol\sigma^-} }
{\sinh{(pu_j^-)}\sinh{(pu_k^-)}}\\
&\phantom{+ p^2 \eta^2\;\;}
+ \frac{1}{2}\sum_j \left(\S_j^z\right)^2 \id 
+ \text{irrelevant terms}
\biggr]
\biggr\}\;,\\
&\text{det}_q T(-u+\eta/2) \simeq 
P^2(-u)\left[1-p\eta \sum_j \coth{(pu_j^+)} \right]\\
&T^{-1}(-u) \simeq ~P^{-1}(-u)\Biggl\{\id +
p\eta\sum_j \frac{1}{\sinh{(pu_j^+)}}
\left[\cosh{(pu_j^+)} 
\S_j^z {\boldsymbol\sigma^z}\!+\!\S_j^+ 
{\boldsymbol\sigma^-}\!+\!\S_j^- {\boldsymbol\sigma^+}\right]
+ \\
&+ p^2 \eta^2 \biggl[\sum_{j<k} 
\frac{\cosh{(pu_j^+)}\cosh{(pu_k^+)} \S_j^z \S_k^z \id + 
\S_j^+ \S_k^- {\boldsymbol\sigma^-} {\boldsymbol\sigma^+} 
+ \S_j^- \S_k^+  {\boldsymbol\sigma^+} {\boldsymbol\sigma^-}}
{\sinh{(pu_j^+)}\sinh{(pu_k^+)}}+\\
&\phantom{+ p^2 \eta^2\;\;} + \frac{1}{2}\sum_j \left(\S_j^z\right)^2 \id  
+ \text{irrelevant terms}
\biggr]
\Biggr\}\;,
\end{align*}
where we defined  $u^\pm_j\equiv u\pm z_j$, 
$P(u)=\prod_j \sinh(p u^-_j)/p$. The irrelevant terms are 
either \CC-numbers or off-diagonal matrices,  
contributing to the trace with terms order $\eta^3$.  
The expansion of $K_\pm$ reads
\begin{equation*}
\begin{split}
K_\pm(u,\xi_\pm) 
\simeq &
\frac{1}{p}
\begin{pmatrix}
\sinh{[p(u+\xi_\pm^{(0)})]} &0 \\
0& -\sinh{[p(u-\xi_\pm^{(0)})]}
\end{pmatrix} + \\
&+ \eta
\begin{pmatrix}
(\delta_{\pm,+}+\xi_\pm^{(1)})\cosh{[p(u+\xi_\pm^{(0)})]} &0 \\
0& -(\delta_{\pm,+}-\xi_\pm^{(1)})\cosh{[p(u-\xi_\pm^{(0)})]}
\end{pmatrix}
\end{split}
\end{equation*}
where $\delta_{\pm,+}$ is the usual Kronecker $\delta$, and 
we took into account the expansion 
\begin{equation}
\label{boundary-expansion}
\xi_\pm \simeq \xi_\pm^{(0)} + \eta\, \xi_\pm^{(1)}\;.
\end{equation}
We define for convenience $2\xi=\xi_+^{(0)} - \xi_-^{(0)}$, 
$2\Sigma=\xi_+^{(0)} + \xi_-^{(0)}$, 
and $2\varsigma=\xi_+^{(1)} + \xi_-^{(1)}$.
Then the terms in the expansion of the transfer matrix given in 
Eq.~(\ref{quasi-transfer}) 
read 
\begin{equation}
\begin{split}
&\hat{\tau}^{(0)}(u)=~ \frac{1}{p^2} P(u) P^{-1}(-u) 
\bigl( \cosh{(2pu)}\cosh{(2p\Sigma)} 
-\cosh{(2p\sigma)}\bigr)
\;,\\
&\hat{\tau}^{(1)}(u)=~\text{\CC-numbers} 
+ \frac{1}{p} P(u) P^{-1}(-u) \times \\
&\times \sinh{(2pu)} \sinh(2p\Sigma) 
\sum_j \bigl(\coth(pu^-_j) + \coth(pu^+_j)\bigr)\S_j^z\;,\\
&\hat{\tau}^{(2)}(u)= \mbox{\CC-numbers} + P(u) P^{-1}(-u)\Biggl\{\\
&2\varsigma \sinh{(2pu)} 
\cosh(2p\Sigma)  
\sum_j \biggl(\coth(pu^-_j) + \coth(pu^+_j)\biggr)\S_j^z + \\
&+ \biggl(\cosh{(2pu)}\sinh{(2p\Sigma)}   
- \sinh{(2p\xi)}\biggr) 
\sum_j \biggl(\coth(pu^-_j) + \coth(pu^+_j)\biggr)\S_j^z + \\
&+\frac{1}{2}\sinh{(2pu)} \sinh(2p\Sigma) 
\sum_{jk} \coth{(pu_j^+)} \coth{(pu_k^+)}\S_j^z +\\
&+\frac{1}{2}\biggl(\cosh{(2pu)}\cosh{(2p\Sigma)}   
- \cosh{(2p\xi)}
\biggr) \times \\ 
&\qquad\times \sum_{jk} \Biggl[ 
\biggl(\coth(pu^-_j) + \coth(pu^+_j)\biggr)
\biggl(\coth(pu^-_k) + \coth(pu^+_k)\biggr) \S_j^z \S_k^z+\\
&\qquad\qquad+\frac{1}{2} \left(
\frac{1}{\sinh(pu^-_j) \sinh(pu^-_k)}\!
+\!\frac{1}{\sinh(pu^+_j) \sinh(pu^+_k)}\right)
\left(\S_{j}^+ \S_k^-\!+\!\S_{j}^- \S_k^+\right)\Biggr]+ \\
&-\frac{1}{2}
\biggl(\cosh{(2pu)}\cosh{(2p\xi)}\!-\!\cosh{(2p\Sigma)} \biggr) 
\sum_{jk}\frac{\S_{j}^+ \S_k^-\!+\!\S_{j}^- \S_k^+}
{\sinh{(pu_j^-)}\sinh{(pu_k^+)}} 
 + \\
&+\frac{1}{2} \sinh{(2pu)}\sinh{(2p\xi)}  
\sum_{jk} \frac{\S_{j}^+ \S_k^-\!-\!\S_{j}^- \S_k^+}
{\sinh{(pu_j^-)}\sinh{(pu_k^+)}} 
\Biggr\}
\end{split}
\end{equation}
As can be seen by Eqs.\eqref{commexp}, the operators 
$\hat{\tau}^{(2)}(u)$ commute with each other if 
\[\comm{\hat{\tau}^{(1)}(u)}{\hat{\tau}^{(3)}(v)}
+\comm{\hat{\tau}^{(3)}(u)}{\hat{\tau}^{(1)}(v)}=0\;.\] 
A sufficient condition for this relation to be fulfilled is that 
$\hat{\tau}^{(1)}(u)$ is just a \CC-number. This requires that 
$\Sigma=0$. 

%%%%%%%%%%%%%%%%%%%%%%%%%%%%%%%%%%%%%%%%%%%%%%%%%%%%%%%%%%%%%%%%%%%%%%%%%%%%

\subsection{Classical boundary}
At first, we assume \emph{classical} boundary. With this term 
we mean that the boundary parameters 
$\xi_\pm$ are assumed to be independent of $\eta$, i.e.~$\varsigma=0$. 
This case was analyzed by one of the authors in Ref.\cite{Hikami95}. 
The commuting family of operators $\hat{\tau}^{(2)}(u)$ given above 
reduces to 
\begin{align*}
&\hat{\tau}_0^{(2)}(u)= 2 P(u) P^{-1}(-u)\sinh^2{(2pu)} \Biggl\{
\sum_j \frac{\S^z}{\cosh{(2pu)}-\cosh{(2pz_j)}} 
\S_j^z + \\ 
&+\sum_{\myfrac{j,k}{j\neq k}} \frac{1}{
\bigl(\cosh(2pu)-\cosh(2pz_j)\bigr)\bigl(\cosh(2pu)-\cosh(2pz_k)\bigr)}
\Biggl[ \\
&\phantom{\mbox{ }+\,}\frac{1}{2}
\biggl(\cosh(2pz_j)+\cosh(2pz_k)-2\cosh(2p\xi)\biggr)\S_j^z \S_k^z + \\
&+\frac{1}{2}\biggl({\cosh[p(z_j+z_k)]-\cosh(2p\xi)\cosh[p(z_j-z_k)]}\biggr)
\left(\S_j^+ \S_k^- +\S_j^- \S_k^+\right) + \\
&-\frac{1}{2}~{\sinh(2p\xi)\sinh[p(z_j-z_k)]}
\left( \S_j^+ \S_k^- - \S_j^- \S_k^+ \right) \Biggr] \Biggr\}\;,
\end{align*}
where we dropped the $\CC$-numbers and the Casimir 
coming from the term $j=k$ in the sums. 
A finite subset of $u$-independent operators in involution can be 
obtained taking the limits $u\to z_j$ of $\hat{\tau}^{(2)}(u)$, and 
dividing by the factor 
\[\sinh(2pz_j) P^{-1}(-z_j)
\prod_{k\neq j} \sinh[p(z_j-z_k)]/p\;.\] 
They are 
\begin{align}
&\hat{\tau}_{0j}= 
\sum_{k} \S_k^z \S_j^z  
+\sum_{\myfrac{k}{k\neq j}} 
\frac{1}{\cosh(2pz_j)-\cosh(2pz_k)}
\Biggl[ \nonumber \\
&\phantom{\mbox{ }+}
\biggl(\cosh(2pz_j)+\cosh(2pz_k)-2\cosh(2p\xi)\biggr)\S_j^z \S_k^z + 
\nonumber \\
&+\biggl({\cosh[p(z_j+z_k)]-\cosh(2p\xi)\cosh[p(z_j-z_k)]}\biggr)
\left(\S_j^+ \S_k^- +\S_j^- \S_k^+\right) + \nonumber \\
&-{\sinh(2p\xi)\sinh[p(z_j-z_k)]}
\left( \S_j^+ \S_k^- - \S_j^- \S_k^+ \right) \Biggr]
\;.
\label{tau-classical}
\end{align}
The $\hat{\tau}_j$ form a complete set, in the sense that any 
$\hat{\tau}^{(2)}(u)$ can be built from them according to the formula 
\begin{equation*}
\hat{\tau}^{(2)}_0(u) = 2~P(u) P^{-1}(-u) \sinh^2{(2pu)}
\sum_j \frac{1}{\cosh(2pu)-\cosh(2pz_j)}~\hat{\tau}_{0j} \;.
\end{equation*}

We notice the  term $\sum_{k} \S_k^z \S_j^z \equiv \S^z \S_j^z$ in the
operators $\hat{\tau}_j$. It  describes a 
self--interaction of the spins with the magnetic field generated 
by the spins themselves. In the next section we shall see 
how to  add an external magnetic field.
\\
The eigenstates of operators~(\ref{tau-classical}) are given by 
\begin{equation}\label{eigenany}
\ket{\Psi}= \prod_{\alpha=1}^{M} \S^-(e_\alpha) \ket{H}\;,
\end{equation}
where \[\S^-(u) =
\sum_j\frac{\cosh[p(u+z_j+2\xi)]-\cosh[p(u-z_j)]}
{\cosh(2pu)-\cosh(2pz_j)} \S_j^- \propto 
\left. \frac{d}{d\eta} {\mathcal B}(u) \right|_{\eta=0}\;.\] 
The rapidities $e_\alpha$ fulfill the first order term in the expansion of 
Eqs.~\eqref{genbethe} around $\eta=0$
\begin{equation}
\begin{split}
&\sum_{\beta\neq\alpha} \frac{1}{\cosh(2pe_\alpha)-\cosh(2pe_\beta)} +\\
&- \sum_j \frac{S_j}{\cosh(2pe_\alpha) - \cosh(2pz_j)} 
+\frac{1/2}{\cosh(2pe_\alpha) - \cosh(2p\xi)} =0\;.
\end{split}
\end{equation}
Putting $\cosh(2pe_\alpha) = \exp(2E_\alpha) + \cosh(2p\xi)$ and 
$\cosh(2pz_j) = \exp(2w_j) + \cosh(2p\xi)$, the equations above reduce to 
the modified Gaudin's equations presented in Refs.~\cite{Amico01a}: 
\begin{equation}
\sum_{\beta\neq\alpha} \coth(E_\alpha-E_\beta) 
- \sum_j S_j \coth(E_\alpha - w_j) + S^z  =0\;.
\end{equation}
The eigenvalues are 
\begin{equation} 
\tau_{0j} = S_j \left(S^z+ \sum_{k\neq j} S_k \coth[(w_j-w_k)]  
- \sum_\alpha \coth{[(w_j-E_\alpha)]}\right) \;.
\end{equation}
%%%%%%%%%%%%%%%%%%%%%%%%%%%%%%%%%%%%%%%%%%%%%%%%%%%%%%%%%%%%%%%%%%%%%%%%%%%%
\subsection{Non-classical boundary}\label{nonclassical}
In this section we show how  to include a scalable 
term proportional to $\S_j^z$ in the operators $\hat{\tau}_j$.
Such a term is crucial for physical applications since, as we shall see, it 
allows to introduce a non-uniform magnetic field in 
the Hamiltonian. Furthermore, when the spins are realized 
by pairs of time-reversed electrons 
\(\S_j^z = - \tfrac{1}{2}(\n_{j\uparrow} + \n_{j\downarrow}-1)\), 
\(\S_j^+ = c_{j\up}c_{j\dwn}\) 
a non-uniform magnetic field corresponds to a kinetic energy 
term (see section~\ref{applications}). 
In order to reach our goal, 
we have exploited the fact that $\xi_\pm$ can depend on $\eta$, i.e.~
\(\xi_+^{(1)} + \xi_-^{(1)}\) is not necessarily zero. 
We refer to this kind of boundary conditions  as a \emph{non-classical} boundary. 
Thus, we put $\varsigma\neq 0$. 
We obtain
\begin{align}
\nonumber
&\hat{\tau}^{(2)}(u)= 2 P(u) P^{-1}(-u) \sinh^2{(2pu)} \Biggl\{
\sum_j 
\frac{2\varsigma+\S^z}{\cosh{(2pu)}-\cosh{(2pz_j)}} 
\S_j^z  + \\ 
\nonumber
&+\sum_{\myfrac{j,k}{j\neq k}} \frac{1}{
\bigl(\cosh(2pu)-\cosh(2pz_j)\bigr)\bigl(\cosh(2pu)-\cosh(2pz_k)\bigr)}
\Biggl[ \\
\nonumber
&\phantom{\mbox{ }+\,}\frac{1}{2}
\biggl(\cosh(2pz_j)+\cosh(2pz_k)-2\cosh(2p\xi)\biggr)\S_j^z \S_k^z + \\
\nonumber
&+\frac{1}{2}\biggl({\cosh[p(z_j+z_k)]-\cosh(2p\xi)\cosh[p(z_j-z_k)]}\biggr)
\left(\S_j^+ \S_k^- +\S_j^- \S_k^+\right) + \\
&-\frac{1}{2}~{\sinh(2p\xi)\sinh[p(z_j-z_k)]}
\left( \S_j^+ \S_k^- - \S_j^- \S_k^+ \right) \Biggr] \Biggr\}\;.
\end{align} 
The integrals of motion are obtained again taking 
the limits $u\to z_j$, 
dividing now also by \(2\varsigma+\S^z\):  
\begin{equation}
\begin{split}
\hat{\tau}_j\doteq \tau_j(\hat{S})=&~\S_j^z  
-J(\S^z)\sum_{\myfrac{k}{k\neq j}} 
\frac{1}{\cosh(2pz_j)-\cosh(2pz_k)}
\Biggl[ \\
&\phantom{\mbox{ }+}
\biggl(\cosh(2pz_j)+\cosh(2pz_k)-2\cosh(2p\xi)\biggr)\S_j^z \S_k^z + \\
&+\biggl({\cosh[p(z_j+z_k)]-\cosh(2p\xi)\cosh[p(z_j-z_k)]}\biggr)
\left(\S_j^+ \S_k^- +\S_j^- \S_k^+\right) + \\
&-{\sinh(2p\xi)\sinh[p(z_j-z_k)]}
\left( \S_j^+ \S_k^- - \S_j^- \S_k^+ \right) \Biggr]
\;,
\end{split}
\label{non-classical-tau}
\end{equation}
where we put $J(\S^z)=J/(1-J\S^z)$, with $J=-1/(2\varsigma)$. 
The operators $\hat{\tau}^{(2)}(u)$ can be built from the $\hat{\tau}_j$ 
according to 
\begin{equation*}
\hat{\tau}^{(2)}(u) = 2(2\varsigma + \S^z) P(u) P^{-1}(-u) \sinh^2{(2pu)}
\sum_j \frac{1}{\cosh(2pu)-\cosh(2pz_j)}~\hat{\tau}_j \;.
\end{equation*}
For real $J$, the $\hat{\tau}_j$ 
are Hermitian if $z_j$ are real and $\xi$ is pure imaginary, 
or vice-versa. 
Their eigenstates are still given by Eq.~\eqref{eigenany}
\begin{equation}
\ket{\Psi}=\prod_{\alpha=1}^{M} \S^-(e_\alpha) \ket{H}\;,
\label{eigenstates}
\end{equation}
where \[\S^-(u) =
\sum_j\frac{\cosh[p(u+z_j+2\xi)]-\cosh[p(u-z_j)]}
{\cosh(2pu)-\cosh(2pz_j)} \S_j^- \propto 
\left. \frac{d}{d\eta} {\mathcal B}(u) \right|_{\eta=0}\;.\] 
The difference with the previous subsection, is that 
the first order term in the expansion of 
Eqs.~\eqref{genbethe} around $\eta=0$ contains an additional term
\begin{equation}\label{betheeq2}
\begin{split}
&\sum_{\beta\neq\alpha} \frac{1}{\cosh(2pe_\alpha)-\cosh(2pe_\beta)} \\
&- \sum_j \frac{S_j}{\cosh(2pe_\alpha) - \cosh(2pz_j)} 
+\frac{\frac{1}{2}(1+{1/J})}{\cosh(2pe_\alpha) - \cosh(2p\xi)} =0\;.
\end{split}
\end{equation}
Putting $\cosh(2pe_\alpha) = \exp(2E_\alpha) + \cosh(2p\xi)$ and 
$\cosh(2pz_j) = \exp(2w_j) + \cosh(2p\xi)$, the equations above reduce as 
well to 
the modified Gaudin's equations presented in Refs.~\cite{Amico01a}: 
\begin{equation}
\sum_{\beta\neq\alpha} \coth(E_\alpha-E_\beta)
- \sum_j S_j \coth(E_\alpha - w_j) + \frac{1}{J(S^z)} =0\;.
\label{quasi-bethe}
\end{equation}
The eigenvalues are 
\begin{equation} 
\tau_j = S_j \left[1 - J(S^z)\left(\sum_{k\neq j} S_k \coth[(w_j-w_k)]  
- \sum_\alpha \coth{[(w_j-E_\alpha)]}\right)\right] \;.
\label{quasi-eigenvalues}
\end{equation}
There are parameterizations yielding \emph{rational} Bethe equations. 
Among these, we consider
\begin{equation}
\frac{1+x_j}{1-x_j} = \cosh(2pz_j) - \cosh(2p\xi)\;\;,\;\;
\frac{1+\la_\alpha}{1-\la_\alpha} = \cosh(2pe_\alpha) - \cosh(2p\xi) \;.
\end{equation}
Then the eigenvalues are
\begin{equation}
\tau_j= 
S_j \left(1  - J(S^z)\sum_{k\neq j} S_k \frac{1 - x_j x_k}{x_j - x_k}
+ J(S^z)\sum_\alpha 
\frac{1 -x_j \la_\alpha}{x_j - \la_\alpha}
\right) \;.
\end{equation}
The Bethe equations \eqref{betheeq2} become 
\begin{equation}
\begin{split}
&\sum_{\beta\neq\alpha} \frac{1}{\la_\alpha-\la_\beta} 
- \sum_j \frac{S_j}{\la_\alpha - x_j} + \\
&+\frac{1}{2J(S^z)}\left(\frac{1+J(S^z)(1+S^z)}{1+\la_\alpha} 
+\frac{1-J(S^z)(1+S^z)}{1-\la_\alpha}\right) =0\;.
\end{split}
\end{equation}
In this form, they admit a two dimensional electrostatic interpretation
\cite{Electro2002}. 

%%%%%%%%%%%%%%%%%%%%%%%%%%%%%%%%%%%%%%%%%%%%%%%%%%%%%%%%%%%%%%%%%%%%%%%%%%%%

\subsection{On the equivalence with  Gaudin's model in external magnetic field}\label{noreflection}

We show that, at $\xi=0$ and $p\xi=i\pi/2$, 
these integrals of motion are equivalent to the 
modified Gaudin's Hamiltonians introduced in Ref.\cite{Amico01a}. 
The integrals of motion reduce to
\begin{equation}
\begin{split}
\hat{\tau}_j =&  \S_j^z - J(\S^z)\sum_{k\neq j}\Biggl\{ 
\frac{\cosh(2pz_j)+\cosh(2pz_k)\mp 2}{\cosh(2pz_j)-\cosh(2pz_k)}\S_j^z \S_k^z 
+ \\
&\phantom{+\sum_{k\neq j}\Biggl\{}
+\frac{\cosh[p(z_j+z_k)]\mp\cosh[p(z_j-z_k)]}
{\cosh(2pz_j)-\cosh(2pz_k)}
\left(\S_j^+ \S_k^- +\S_j^- \S_k^+ \right)\Biggr\}
\;,
\end{split}
\end{equation}
where the upper sign refers to $\xi=0$ and the lower one to $p\xi=i\pi/2$. 
We make the change of variable  \(\sinh{(pz_j)}= \exp{w_j}\), if $\xi=0$, 
\(\cosh{(pz_j)}= \exp{w_j}\), if  $p\xi=i\pi/2$, 
obtaining 
\begin{equation*}
\hat{\tau}_j = \S_j^z - J(\S^z)\sum_{k\neq j}\Biggl\{ 
\coth(w_j-w_k) \S_j^z \S_k^z 
+\frac{1}{2\sinh(w_j-w_k)}
\left(\S_j^+ \S_k^- +\S_j^- \S_k^+ \right)\Biggr\}
\;.
\end{equation*}
Thus, apart a sector-dependent rescaling of the coupling 
\(J\rightarrow J(\S^z)=J~(1-J \S^z)^{-1}\), the 
operators given above are equivalent to modified Gaudin's Hamiltonians. 

%%%%%%%%%%%%%%%%%%%%%%%%%%%%%%%%%%%%%%%%%%%%%%%%%%%%%%%%%%%%%%%%%%%%%%%%%%%%

\subsection{Construction of the  Hamiltonian}
\label{H-construction}
The simplest Hamiltonian that is possible to be built up 
is a first degree polynomial in $\hat{\tau}_j$ with arbitrary 
real parameters  $h_j$ 
\begin{equation}
\label{hamiltonian}
\begin{split}
H=&\sum_j 2 h_j~\hat{\tau}_j = \sum_j 2 h_j~\S_j^z 
-J(\S^z)\sum_{\myfrac{j,k}{j\neq k}}
\frac{h_j-h_k}{\cosh(2pz_j)-\cosh(2pz_k)}
\Biggl[ \\
&\phantom{\mbox{ }+\,\;}
{\bigl(\cosh(2pz_j)+\cosh(2pz_k)-2\cos{(2pt)}\bigr)}\S_j^z \S_k^z + \\
&\mbox{ }+\bigl({\cosh[p(z_j+z_k)]\!-\!\cos(2pt)\cosh[p(z_j-z_k)]}\bigr)
\left(\S_j^+ \S_k^-\!+\!\S_j^- \S_k^+\right) + \\
&\phantom{1234567890123456}\mbox{ }-{i\sin(2pt)\sinh[p(z_j-z_k)]}
\left( \S_j^+ \S_k^- -\S_j^- \S_k^+ \right) \Biggr]
\;,
\end{split}
\end{equation}
where we put $\xi=it$, with real $t$.

%%%%%%%%%%%%%%%%%%%%%%%%%%%%%%%%%%%%%%%%%%%%%%%%%%%%%%%%%%%%%%%%%%%%%%%%%%%%

\section{Second quantized Hamiltonians}\label{applications}
In this section we employ the fermionic realization of $su(2)$
to write the integrable models we found Eq.~(\ref{hamiltonian}) 
in second quantization. The 
two orthogonal $D_j$ and $D'_l$-dimensional realizations are
\begin{equation}
\K_j^+ =  \sum_{\delta_j=1}^{D_j}  
c_{j,\delta_j\dwn} c_{j\delta_j\up},\quad \quad 
\K_j^- = \left(\K_j^+\right)^\d , \qquad 
\K_j^z = \frac{1}{2} \sum_{\delta_j =1}^{D_j}
\left(1- \n_{j\delta_j\up} - \n_{j\delta_j\dwn}\right) ,
\label{K-spin}
\end{equation}
and
\begin{equation}
\S_l^+ = \sum_{\rho_l=1}^{D'_l} c^\d_{l\rho_l\up} c^{}_{l\rho_l\dwn} 
\quad, \quad 
\S_l^- = \left(\S_l^+\right)^\d \quad, \quad
\S_l^z = \frac{1}{2} \sum_{\rho_l=1}^{D'_l} 
\left(\n_{l\rho_l\up} - \n_{l\rho_l\dwn}\right), 
\label{S-spin}
\end{equation}
where operators $c$, $c^\d$, and $n\equiv c^\d c$ are fermionic operators. 
We arbitrarily grouped the levels in the subsets 
$j=1,\dotsc,\Omega_K$, each containing $D_j$ levels, and in the subsets
$l=1,\dotsc,\Omega_S$, each containing $D'_l$ levels. 
The maximum values of the $z$ components of the spin are 
$K_j = D_j/2$ and $S_l = D'_l/2$ respectively. 
Thus, a level $a$ will be characterized alternatively 
by the pairs $(j(a),\delta_{j(a)}(a))$ or 
$(l(a),\rho_{l(a)}(a))$. 
We write a Hamiltonian of the form
\begin{equation}
\label{secondq-Hamiltonian}
H=H_K+H_S+E_0\;,
\end{equation}
where $E_0$ is a constant and 
\begin{equation}
\begin{split}
H_K=&-\sum_{j=1}^{\Omega_K} 
2\eta_j \tau_j(\hat{K}) 
+\sum_{j=1}^{\Omega_K}  g_{jj} \boldsymbol{\K}_j^2 
\;,
\end{split}
\end{equation}
\begin{equation}
\begin{split}
H_S=&\sum_{l=1}^{\Omega_S} 2\zeta_l \tau_l(\hat{S}) 
+ \sum_{l=1}^{\Omega_S} J^{xx}_{ll} \boldsymbol{\S}_l^2
\;,
\end{split}
\end{equation}
where operators $\tau(\hat{O})\, ,\; \hat{O}=\K,\S$ are defined in 
Eq.~(\ref{non-classical-tau}). Due to the orthogonality of the 
realizations~(\ref{K-spin}), (\ref{S-spin}) we observe that 
$\left[H_K,H_S\right ]=0$. 
Furthermore, $H_K$ and $H_S$ are block-diagonal, and their 
common eigenstates are the direct product of the eigenstates 
of $H_K$ and of $H_S$, each restricted 
to the subspace corresponding to one of its blocks \cite{Amico01b}. 
The integrability together 
with the exact solution of the Hamiltonian 
(\ref{secondq-Hamiltonian}) follows from the integrability 
of each $H_K$, $H_S$ proved in Section~\ref{nonclassical}
and from Eqs.~(\ref{eigenstates}),~(\ref{quasi-bethe}), 
and (\ref{quasi-eigenvalues}). 

Finally, the second quantized form of the Hamiltonian 
\eqref{secondq-Hamiltonian} reads
\begin{equation}\label{ham}
\begin{split}
H = \sum_{a\sigma} \eps_{a\sigma} \n_{a\sigma} + 
\sum_{ab} \biggl[&
U_{ab} (\n_{a\up}+\n_{a\dwn})  (\n_{b\up}+\n_{b\dwn})+
g_{ab} c_{a\up}^\dag c_{a\dwn}^\dag c_{b\dwn}c_{b\up} + 
 \\
&+J^z_{ab}  (\n_{a\up}-\n_{a\dwn})  (\n_{b\up}-\n_{b\dwn})+
J^{xx}_{ab} c_{a\up}^\dag c_{b\dwn}^\dag c_{b\up}c_{a\dwn}
\biggr]\;,
\end{split}
\end{equation}
where  $\Omega$ number the levels,  
$a,b=1,\dotsc,\Omega$ and 
$c_{a,\sigma}\equiv c_{j(a),\delta_{j(a)}(a),\sigma}$; the constant in 
Eq.~(\ref{secondq-Hamiltonian}) turns out to be 
$E_0=\sum_j D_j\, \eps_j+\sum_{jk}D_jD_k U_{jk}$.\\
The kinetic energy term reads 
\[\sum_{a\sigma} \eps_{a\sigma} \n_{a\sigma} = 
\sum_{a} \left[\frac{1}{2} (\eps_{a\up} + \eps_{a\dwn}) 
(\n_{a\up} + \n_{a\dwn}) + \frac{1}{2} (\eps_{a\up} - \eps_{a\dwn}) 
(\n_{a\up} - \n_{a\dwn}) \right]\,.\]   
We choose a partition ---in equivalence classes--- 
of the single particle levels 
in such a way that all levels having the same value of 
\(\eps_a \equiv \frac{1}{2} (\eps_{a\up} + \eps_{a\dwn})\) belong 
to the same class (hence we write $\eps_j$ instead of 
$\eps_a$, where $j$ individuates the class)~\footnote{$\eta_j=\eps_j  + 
2\sum_k D_k U_{jk} +g_{jj}/2$ have to be determined 
consistently; they must satisfy a system of linear equations, 
as discussed in \cite{Dilorenzo01}.}. Analogously, a 
second partition is defined in such a way that all 
the levels having the same value of 
\(\zeta_a \equiv \frac{1}{2} (\eps_{a\up} - \eps_{a\dwn})\) belong 
to the same class (hence we write the common value as $\zeta_l$)
\footnote{The partitioning of 
the levels that we chose above guarantee that the interaction does 
not vanish between levels having the same value of $\eps_a$ or 
$\zeta_a$. 
}. 
The couplings between levels $a$ and $b$ 
depend only on the equivalence classes of the two levels. 
For $j\equiv j(a)\neq k\equiv k(b)$ and $l\equiv l(a)\neq m\equiv m(b)$, 
they are
\begin{align}
\begin{aligned}
g_{ab} &= g_{jk} = 2 J_K(K^z){\left(\eta_{j}-\eta_{k}\right)}
\frac{{\cosh[p(z_j+z_k)]\!-\!\cosh[p(z_j-z_k-2it_K)]}}
{\cosh(2pz_j)-\cosh(2pz_k)}\;,\\
4 U_{ab} &= 4 U_{jk} = J_K(K^z) {(\eta_j-\eta_k)}
\frac{\cosh(2pz_j)+\cosh(2pz_k)-2\cos{(2pt_K)}}
{\cosh(2pz_j)-\cosh(2pz_k)}\;,\\
J^{xx}_{ab}&= J^{xx}_{lm}= -2J_S(S^z){(\zeta_l-\zeta_m)}
\frac{{\cosh[p'(y_l+y_m)]\!-\!\cosh[p'(y_l-y_m+2it_S)]}}
{\cosh(2p'y_l)-\cosh(2p'y_m)}\;,\\
J^z_{ab}&= J^z_{lm}= -J_S(S^z){(\zeta_l-\zeta_m)}
\frac{\cosh(2p'y_l)+\cosh(2p'y_m)-2\cos{(2p't_S)}}
{\cosh(2p'y_l)-\cosh(2p'y_m)}\;.
\end{aligned} 
\label{fermionic-couplings}
\end{align}
For $j=k$, we have the relation $g_{jj}=4U_{jj}$, and  
$g_{jj}$ can be chosen arbitrarily.\footnote{In particular, they 
can be chosen in such a way that $\eta_j=\eps_j$~\cite{Richardson-private}.} 
Analogously, for $l=m$, we have $J^z_{ll}=J^{xx}_{ll}$.

%%%%%%%%%%%%%%%%%%%%%%%%%%%%%%%%%%%%%%%%%%%%%%%%%%%%%%%%%%%%%%%%%%%%%%%%%%%%
\section{Conclusions}\label{concls}

In this paper we have studied integrable disordered vertex models in presence 
of boundary reflections. The quasi-classical expansion of the models 
has been thoroughly investigated. This expansion produces a hierarchy 
of models which import the integrability of the original vertex models.
The extraction of the energy from the generating functionals 
(which is unfeasible, in general) is very simplified by 
the quasi-classical limit and the Hamiltonian is given as polynomial of the 
integrals of motion. The class of models we obtain describes 
interacting spins 
with non uniform couplings, and in a non uniform 
external magnetic field.  In this sense the present models generalize 
those ones found in Ref.~\cite{Hikami95}. On the other hand, these 
Hamiltonians constitutes also a one-parameter ($t$ in the text) 
extension of the class of models found in 
Ref.~\cite{Amico01a} that are recovered when the boundary terms give rise of  
twisted periodic boundary conditions. As a result, the integrability 
of these latter  models has a firm ground within the Sklyanin 
procedure~\cite{Sklyanin88}.    
The presence of the external magnetic field is 
an effect of the boundary terms which are  assumed, in turn, quasi-classical.
We also obtained the exact solution of the class of models presented 
here through Algebraic Bethe Ansatz. 
The Bethe equations can be recast in a form which allows the electrostatic 
analogy as was done in Ref.\cite{Electro2002}. 

An important point is that the models apply to any spin $S_j$ (not only 
to spin $1/2$). 
The reason is that, in the present  case, 
the integrals of motion can contain only spin $S$ operators 
since the quantum algebra $su_q(2)$ reduces to $su(2)$ 
in the quasi-classical limit (whatever the dimension of the 
representation is). 

By realizing the spin operators in terms of fermions, the class of models 
we found describes confined fermions in degenerate levels 
with pairing force interaction.  
\appendix

%%%%%%%%%%%%%%%%%%%%%%%%%%%%%%%%%%%%%%%%%%%%%%%%%%%%%%%%%%%%%%%%%%%%%%%%%%%%

\section{Inhomogeneous vertex models}\label{VM}
\begin{figure}[htb]
\centering{
\resizebox{0.75\textwidth}{!}{%
\subfigure{\includegraphics{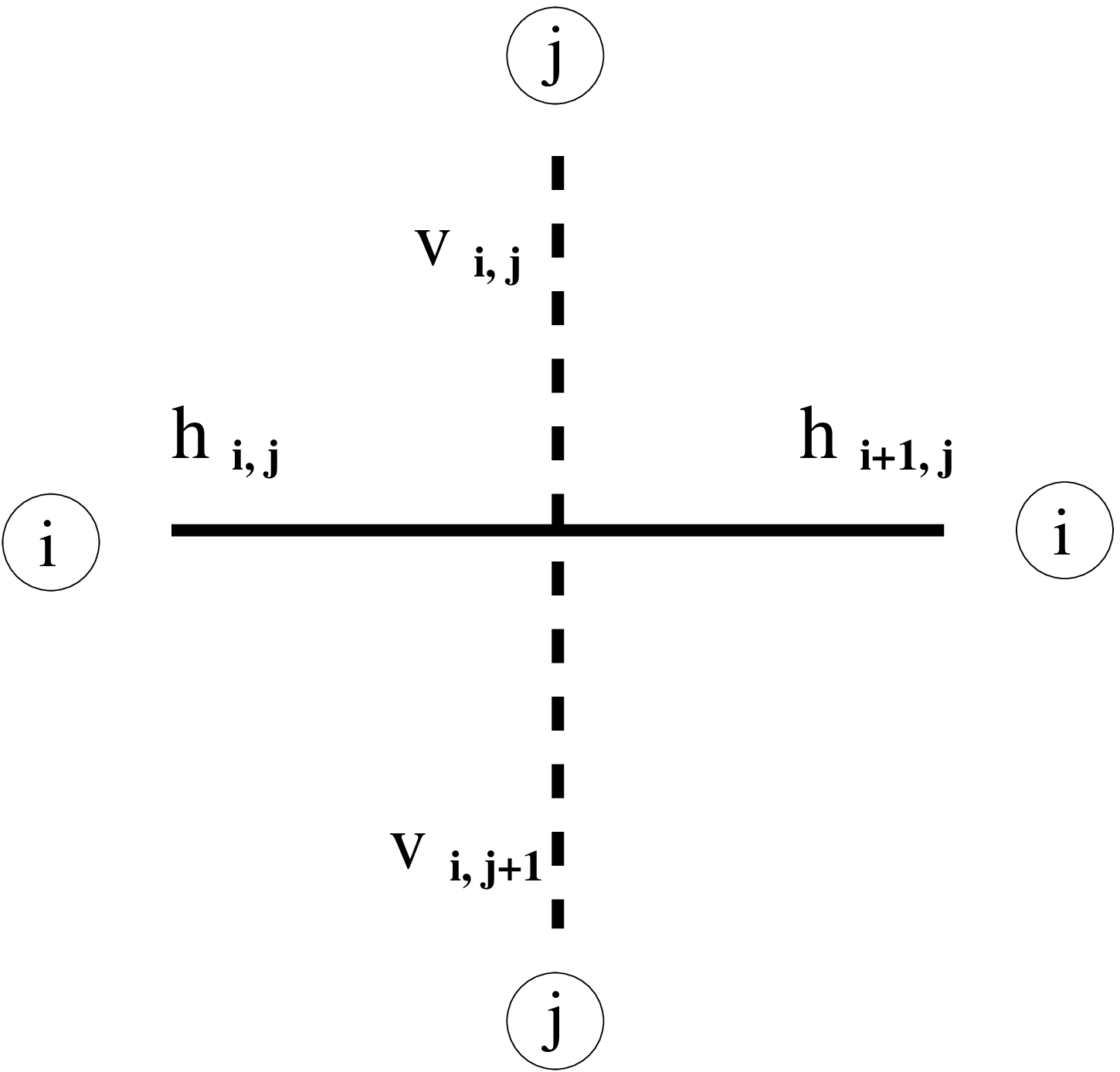}}
\subfigure{\includegraphics{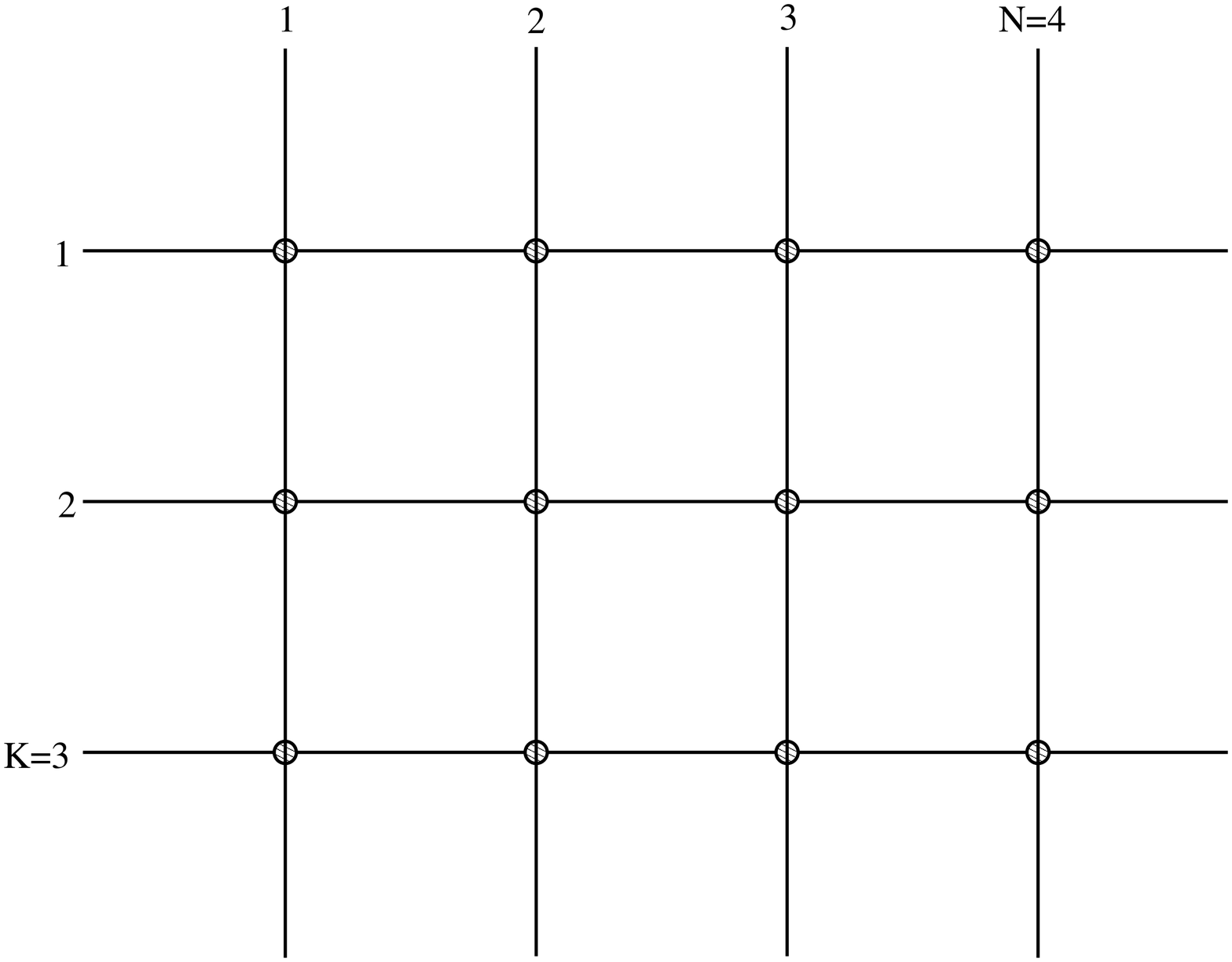}
}}}
\caption{(a) A vertex configuration. (b) Numbering of the lattice.
} 
\label{fig:lattice}
\end{figure}

VM are  models of $2D$ classical statistical mechanics. 
They consist in a $(K\times N)$ array of vertices 
(see Fig.\ref{fig:lattice}), where the nearest 
neighbours are linked by horizontal and vertical legs. 
The legs can be of several species, each identified by a number 
$h_{i,j}=1,\cdots, H_i$, for the horizontal legs of the $i$-th row, 
and $v_{i,j}=1,\cdots, V_j$ 
for the vertical ones of the $j$-th 
column (the number of species can depend on the row or column; in this case, 
we have an \emph{inhomogeneous} model). 
Here we are using the convention that the pair $(i,j)$ individuates 
the vertical (horizontal) leg above (left of) the vertex $(i,j)$. 
A statistical weight $w(legs_{i,j};i,j)=\exp{[-\beta~\eps(legs_{i,j};i,j)]}$ 
is assigned to each vertex, depending on the 
legs configurations (\(legs_{i,j}=\{h_{i,j},h_{i+1,j},v_{i,j},v_{i,j+1}\}\)) 
around it. If the weight depends explicitly on the position 
of the vertex, we have a \emph{disordered} model. 
The goal is to find the partition function 
\(Z = \sum_{\{legs\}} \prod_{i,j} w(legs_{i,j};i,j)\). 

To each vertex, one can associate a matrix, whose elements are 
the weights corresponding to the possible configurations of legs, in the 
following way: 
fix the horizontal legs around site $(i,j)$ to their minimal values, say 
$h_{i,j}=h_{i+1,j}=1$; then 
vary the values of the vertical legs, associating the upper one 
to a row, and the lower one to a column; a $(V_j\times V_j)$ matrix 
is thus obtained, which we indicate by $L_j^{1,1}(i)$, 
whose entries are the weights corresponding to the 
possible legs configurations with horizontal legs fixed to 1; 
then repeat the procedure changing the values of 
the horizontal legs, associating the left one to a row, and the 
right one to a column; a block matrix $L_j(i)$ 
is finally obtained, which is 
conventionally called the Lax operator. It is a $(H_i \times H_i)$ 
matrix whose entries $L_j^{h_{i,j},h_{i+1,j}}(i)$ 
are in turn $(V_j\times V_j)$ matrices, i.e.~operators over the 
linear space ${\mathcal H}_j$. 
The partition function of the $(1\times N)$ lattice with periodic 
boundary conditions in vertical and horizontal direction is, 
in terms of the Lax operators, 
\(Z_1 = \text{Tr}_V \text{Tr}_H\{L_1(1) \dotsm L_N(1)\}\equiv 
\text{Tr}_V \hat{t}(1)\), where by $\text{Tr}_H$ 
we mean the trace over the horizontal space, and by 
$\text{Tr}_V$ the trace over the vertical ones; we introduced the 
transfer matrix \(
\hat{t}(i)\equiv \text{Tr}_H\{L_1(i) \dotsm L_N(i)\}\), where the hat 
is meant to remind that the transfer matrix is an operator over 
$\mathcal H = \bigotimes_j {\mathcal H}_j $, 
the direct product of the 
linear spaces associated to the vertical legs. 
For a $(K\times N)$ lattice, the partition function is 
\(Z = \text{Tr}_V \{\hat{t}(1) \dots \hat{t}(K)\}\). 
If $[\hat{t}(i),\hat{t}(i')]=0$ $\forall i,i'$, 
it is possible to simultaneously diagonalize the $\hat{t}(i)$, 
obtaining \(Z = \sum_r \prod_{i=1}^K t_r(i)\), where 
$t_r(i)$ is the $r$-th eigenvalue of $\hat{t}(i)$. 
In this case, the VM is exactly solvable. \\
From the $\hat{t}(i)$, it 
is commonly possible to extract many-body Hamiltonians of interest.\footnote{
In general, such Hamiltonians are not by any means 
related to the Hamiltonian of the VM.}
 Thus, a given exactly solvable vertex model corresponds uniquely to a family 
of commuting many-body operators.\\ 
It turns out that the transfer matrices commute with each other, and thus the 
corresponding vertex model is exactly solvable, if $H_i = H_{i'}=H$, 
$\forall i,i'$, and a family of 
$(H^2 \times H^2)$ matrices, the $\check{R}$-matrices,  
exists, such that the Lax operators 
obey the relation 
\begin{equation}\label{fund}
\check{R}(i,i')~L_j(i) \otimes L_j(i')
= L_j(i')\otimes L_j(i)~\check{R}(i,i')\;. 
\end{equation}
Given a $\check{R}$-matrix, 
this is a very strict requirement, which in general implies that 
many legs configurations are not allowed, i.e.~their weight is zero, 
while the allowed ones are related to each other by some parametrization. 

A relevant case is when the dimensions of vertical and horizontal space 
are equal, and they do not depend on the row or column: 
$H_i = V_j \equiv 2S +1$. 
Then, the $\check{R}$-matrices are but the Lax operators where the 
matrix elements have been written down explicitly in their matrix 
representation. It turns out that the entries of 
the Lax operators are matrices belonging 
to the $(2S+1)$-dimensional realization of $su(2)$, i.e.~spins over 
the \emph{Hilbert}\footnote{It is a finite-dimensional vector space. We denote it
 as \emph{Hilbert space} in foresight of its interpretation as a quantum space.} 
space ${\mathcal H}_j$. There is the drawback that the 
$\check{R}$-matrix is difficult to determine and to handle, since its 
dimension increases very fast with $S$~\cite{Fateev80,Sogo83}. A technique to build larger 
 $\check{R}$-matrices using the 
$(4\times 4)$ $\check{R}$-matrices (the simplest ones) as 
building blocks was devised by 
Kulish, Reshetikhin, and Sklyanin~\cite{Kulish81}. 
In the present paper, by means of the quasi-classical expansion, we will build 
up operators for spins higher than $1/2$ still making use of $(4\times 4)$ 
$\check{R}$-matrices.

%%%
\section{General $K_+$ matrix}\label{genrefl}
In this appendix we construct the Hamiltonian when the general 
solution of the
reflection equation \eqref{reflplus} is 
considered~\cite{DeVega93,Ghoshal94}.  In this case we have
\begin{equation}
K_+(u)=\frac{1}{p}\left(\begin{matrix}
\sinh[p(u+\eta+\xi_+)]\quad&\kappa_+ \sinh[2p(u+\eta)]\\
\kappa_+ \sinh[2p(u+\eta)]\quad&-\sinh[p(u+\eta-\xi_+)]\quad
\end{matrix}\right)\;.
\label{off-diagonal-K}
\end{equation} 
Since we want $\hat{\tau}^{(1)}$ to be a \CC-number, we must impose  
$\kappa_+ \simeq i \eta c$. 
Thus, the final effect of the general reflection results in an additional 
term in the second order of the transfer matrix
\begin{equation}
\begin{split}
&\hat{\tau}^{(2)}(u) \to \hat{\tau}^{(2)}(u) 
+  \frac{2i c }{p} P(u)P^{-1}(-u)\sinh^2{(2pu)}  
\Biggl\{\\
&\sum_j 
\frac{\sinh[p(z_j-\xi)]}{\cosh(2pu)-\cosh(2pz_j)}\S_j^+ 
-\sum_j 
\frac{\sinh[p(z_j+\xi)]}{\cosh(2pu)-\cosh(2pz_j)}
\S_j^- \Biggr\} 
\;.
\end{split}
\end{equation}
The Hamiltonian is again built according to 
\begin{equation}
H=\sum_j 2 h_j \hat{\tau}_j
\label{off-diagonal-ham}
\end{equation}
where the integrals of motion are  
\begin{equation}
\begin{split}
\hat{\tau}_j =&~(1-J\,\S^z)\S_j^z  
-ic\,J\left(
\frac{\sinh[p(z_j -\xi)]}{p} \S_j^+ - 
\frac{\sinh[p(z_j +\xi)]}{p} \S_j^-\right)  + \\
&-J \sum_{\myfrac{k}{k\neq j}} 
\frac{1}{\cosh(2pz_j)-\cosh(2pz_k)}
\Biggl[ \\
&\phantom{\mbox{ }+}
\bigl(\cosh(2pz_j)+\cosh(2pz_k)-2\cosh(2p\xi)\bigr)\S_j^z \S_k^z + \\
&+\bigl({\cosh[p(z_j+z_k)]-\cosh(2p\xi)\cosh[p(z_j-z_k)]}\bigr)
\left(\S_j^+ \S_k^- +\S_j^- \S_k^+\right) + \\
&-{\sinh(2p\xi)\sinh[p(z_j-z_k)]}
\left( \S_j^+ \S_k^- - \S_j^- \S_k^+ \right) \Biggr]
\; .
\end{split}
\label{off-diagonal-tau}
\end{equation}
We point out that the Hamiltonian is  hermitian for real $c$ and $\xi=it$ with real $t$. 
The diagonalization of this class of Hamiltonians might be achieved by 
functional Bethe ansatz\cite{Sklyanin89}. Nevertheless,
it seems worth to study the models (\ref{off-diagonal-ham}), 
(\ref{off-diagonal-tau})  since their potential  
application  to condensed matter (see also Eqs~(\ref{K-spin}), (\ref{S-spin})).

%%%%%%%%%%%%%%%%%%%%%%%%%%%%
\section{Generalization to $su(n)$}
\label{sec:sun}

In this appendix we briefly discuss on a generalization of the Gaudin
model to the $su(n)$ case.
We can define the Gaudin model for other Lie algebras
(see, e.g., Ref.~\cite{Kulish01} for recent works). 
Following the method depicted in section~\ref{integrability} we obtain a 
modified Gaudin Hamiltonian
to include a scalable term proportional to  the Cartan 
generators of $su(n)$ (as far as we know, previously obtained models 
do not contain this term).

The trigonometric  $R$-matrix for $su(n)$ chains is given by
\begin{multline}
  R(\lambda)
  =
  \frac{1}{p}
  \sum_{a=1}^n \sinh(p (\lambda+\eta)) E^{aa} \otimes E^{aa}
  +
  \frac{1}{p}
  \sum_{a\neq b}^n \sinh(p \lambda) E^{aa} \otimes E^{bb}
  \\
  +
  \frac{1}{p}
  \sum_{a \neq b}^n
  e^{-p \lambda {\rm sgn}(a-b)} \sinh(p \eta) 
  E^{ab}\otimes E^{ba} ,
\end{multline}
where $E^{ab}$ denotes $n\times n$ matrix with unity at $(a,b)$ element.
They satisfy
\begin{equation*}
  [\Hat{E}^{ab} , \Hat{E}^{cd}]
  =
  \Bigl(
  \delta^{bc}
  \Hat{E}^{ad}
  -
  \delta^{da}
  \Hat{E}^{cb}
  \Bigr) .
\end{equation*}
The corresponding  diagonal solution of the 
reflection equation~\eqref{reflminus} is~\cite{DeVega93}
\begin{align}
  K_-(\lambda)
  & =
  \frac{1}{p}
  \sum_{a=1}^{\ell_-}
  e^{p \lambda} \sinh(p (\xi_- - \lambda)) E^{aa}
  +
  \frac{1}{p}
  \sum_{a=\ell_-+1}^n
  e^{-p \lambda} \sinh(p (\xi_-  +\lambda)) E^{aa} ,
  \nonumber \\
  K_+(\lambda)
  & =
  \frac{1}{p}
  \sum_{a=1}^{\ell_+}
  e^{-p\lambda-2p \eta a} \sinh(p(\xi_++\lambda)) E^{aa}
  \nonumber
  \\
  & \qquad
  +
  \frac{1}{p}
  \sum_{a=\ell_+ +1}^n
  e^{p \lambda + p \eta (n-2 a)}
  \sinh(p(\xi_+- \eta n - \lambda))
  E^{aa}
\end{align}
where $\ell_\pm$ is arbitrary, $1 \leq \ell_\pm \leq n$.
Hereafter we set $\ell_+ = \ell_- = \ell$ for simplicity.
With these $K$-matrices,
the Hamiltonian of the $su(n)$ homogenous spin chain with 
nearest neighbour interaction 
with open boundary was computed in  Refs.~\cite{DeVega94,Doikou98} 
by using the formula \eqref{log}. 
The Hamiltonian with the long range interaction is 
constructed following the procedure presented 
in section \ref{H-construction}, where the constants of motion are calculated 
by the formula \eqref{quasi-transfer}. They read
\begin{multline}
  \Hat{\tau}_j
  =
  2 \sum_{a=1}^n a
  \Hat{E}_{j}^{aa}
  \\
  +
  \sum_{a=1}^\ell
  \biggl[
  \xi_-^{(1)} \coth(p (\xi-z_j))
  +
  \xi_+^{(1)} \coth(p (\xi+z_j))
  +
  (n-\ell)
  \frac{\sinh(2 p z_j)}{\sinh(p(\xi-z_j)) \sinh(p(\xi+z_j))}
  \biggr]
  \Hat{E}_j^{aa}
  \\
  +
  \sum_{a=\ell+1}^n
  \biggl[
  \xi_-^{(1)} \coth(p (\xi+z_j))
  +
  \xi_+^{(1)} \coth(p (\xi-z_j))
  \biggr]
  \Hat{E}_j^{aa}
  \\
  +
  \sum_{k \neq j}
  \Biggl[
  \bigl(
  \coth(p(z_j-z_k))+ \coth(p(z_j+z_k))
  \bigr) \sum_a^n \Hat{E}_j^{aa}
% \otimes
  \Hat{E}_{k}^{aa}
  \\
  +
  \sum_{a \neq b}^n
  \frac{e^{p(z_j-z_k) {\rm sgn}(a-b)}}{\sinh(p (z_j-z_k))}
  \Hat{E}_{j}^{ba}
% \otimes
  \Hat{E}_{k}^{ab}
  +
  \sum_{a=1}^\ell \sum_{b=\ell+1}^n
  \frac{\sinh(p(\xi+z_j))}{\sinh(p(\xi-z_j))}
  \frac{e^{p (z_k-z_j)}}{\sinh(p(z_j+z_k))}
  \Hat{E}_{j}^{ba}
% \otimes
  \Hat{E}_{k}^{ab}
  \\
  +
  \sum_{a=\ell+1}^n \sum_{b=1}^\ell
  \frac{\sinh(p(\xi-z_j))}{\sinh(p(\xi+z_j))}
  \frac{e^{p (z_j-z_k)}}{\sinh(p(z_j+z_k))}
  \Hat{E}_{j}^{ba}
% \otimes
  \Hat{E}_{k}^{ab}
  \\
  +
  \sum_{\substack{
      a,b=1
      \\
      a \neq b}}^\ell
  \frac{e^{p (z_j+z_k) {\rm sgn}(b-a)}}{\sinh(p(z_j+z_k))}
  \Hat{E}_{j}^{ba}
%  \otimes
  \Hat{E}_{k}^{ab}
  +
  \sum_{\substack{
      a,b=\ell+1
      \\
      a \neq b}}^n
  \frac{e^{p (z_j+z_k) {\rm sgn}(b-a)}}{\sinh(p(z_j+z_k))}
  \Hat{E}_{j}^{ba} 
%\otimes
  \Hat{E}_{k}^{ab}
  \Biggr]
\end{multline}
where $\Hat{E}_{k}^{ab}$ are site-$k$ $su(n)$ operators. 

The spectrum  of 
$\Hat{\tau}_j$ is given as  limit of 
the eigenvalues of the $su(n)$-transfer matrix (Eq.(6) of~\cite{DeVega94}) and with $u\to z_j$: 
\begin{multline}
  \tau_j
  =
  \xi_+^{(1)} \,
  \Bigl(\coth[p(z_j+\xi)] - \coth(p\xi)\Bigr)
  -
  \xi_-^{(1)} \,
  \Bigl(\coth[p(z_j+\xi)] + \coth(p\xi)\Bigr)
  \\
  +
  (\ell-n)
  \frac{\sinh(2 p z_j)}{
    \sinh[p(z_j - \xi)] \sinh[p(z_j+\xi)]
  }
  + 1 - n \frac{e^{-2 z_j}}{\sinh(2 z_j)}
  \\
  +
  \sum_{k \neq j}^N
  \Bigl(
  \coth[p (z_j+z_k)]
  +
  \coth[p ( z_j-z_k)]
  \Bigr)
  \\
  \hspace{5cm}+ 
  \sum_{k}^{M_1}
  \Bigl(
  \coth[p (z_j + e_k^{(1)})]
  +
  \coth[p ( z_j-e_k^{(1)})]
  \Bigr)
\end{multline}
The Bethe ansatz equations can be obtained in the same  limit of a
result in Ref.~\cite{DeVega94} , and we have
(for $a=1,2, \dots,n-1$)
\begin{multline}
  2\sum_{k \neq j}^{M_a}
  \Bigl(
  \coth[p(e_j^{(a)} + e_k^{(a)})]
  +
  \coth[p(e_j^{(a)} - e_k^{(a)})]
  \Bigr)
  \\
  +
  \delta_{a,\ell} \,
  \Bigl(
  n+\xi_-^{(1)} - \xi_+^{(1)}
  \Bigr)  \,
  \Bigl(
  \coth[p(z-\xi)]
  +
  \coth[p(z+\xi)]
  \Bigr)
  \\
  =
  \sum_{k}^{M_{a+1}}
  \Bigl(
  \coth[p(e_j^{(a)}+e_k^{(a+1)})]
  +
  \coth[p(e_j^{(a)}-e_k^{(a+1)})]
  \Bigr)
  \\
  +
  \sum_{k}^{M_{a-1}}
  \Bigl(
  \coth[p(e_j^{(a)}+e_k^{(a-1)})]
  +
  \coth[p(e_j^{(a)}-e_k^{(a-1)})]
  \Bigr)
\end{multline}
Here we assume $e_j^{(0)}=z_j$ and $M_0=N$,
$M_n=0$.

The fermionic models can be obtained  by using the fermionic realization 
\begin{equation}
  \Hat{E}^{ab}_j
  =
  c_{j,a}^\dagger \, c_{j,b} - \frac{1}{n} \, \delta_{ab} ,
\end{equation}
with a constraint
\begin{equation*}
  \sum_a c_{j,a}^\dagger \, c_{j,a} = 1.
\end{equation*}

%%%%%%%%%%%%%%%%%%%%%%%%%%%%%%%%%%%%%%%%%%%%%%%%%%%%%%%%%%%%%%%%%%%%%%%%%%

%%%%%%%%%%%%%%%%%%%%%%%%%%%%%%%%%%%%%%%%%%%%%%%%%%%%%%%%%%%%


\begin{thebibliography}{10}

\bibitem{Baxter82}
R.~J. Baxter.
\newblock {\em Exactly Solved Models in Statistical Mechanics}.
\newblock Academic Press (1982).

\bibitem{Gaudin83}
M. Gaudin.
\newblock {\em La fonction d'onde de Bethe}.
\newblock Masson (1983).

\bibitem{Korepin93}
V.~E. Korepin, N.~M. Bogoliubov, and A.~G. Izergin.
\newblock {\em Quantum Inverse Scattering Method and Correlation Functions}.
\newblock Cambridge Univ.~Press (1993).

\bibitem{Gaudin76}
M.~Gaudin, 
  J. Physique {\bf 37}, 1087 (1976).

\bibitem{Hikami92}
K.~Hikami, P.~P. Kulish, and M.~Wadati, 
J.~Phys.~Soc.~Jpn {\bf 61}, 3071 (1992).

\bibitem{Sklyanin89}
E.~K. Sklyanin,   J.~Sov.~Math. {\bf 47}, 2473 (1989).

\bibitem{Cambiaggio97}
M.~C. Cambiaggio, A.~M.~F. Rivas, and M.~Saraceno, 
Nucl.~Phys.~A {\bf 624}, 157 (1997).

\bibitem{Richardson63a}
R.~W. Richardson, 
Phys.~Lett. {\bf 3}, 277 (1963).

\bibitem{RalphvonDelft01}
J.~von Delft and D.~C. Ralph, 
Physics Reports {\bf 345}, 61 (2001).

\bibitem{Mastellone98}
A.~Mastellone, G.~Falci, and R.~Fazio, 
Phys.~Rev.~Lett. {\bf 80}, 4542 (1998).

\bibitem{Amico01}
L. Amico, G.~Falci, and R. Fazio, 
J.~Phys.~A {\bf 34}, 6425--6434 (2001).

\bibitem{Sierra00}
G. Sierra, 
Nucl.~Phys.~B {\bf 572}, 517--534 (2000).

\bibitem{Babujian93}
H.~M. Babujian, J. Phys. A {\bf 26}, 6981 (1993).

\bibitem{Reshetikhin95}
N.~Reshetikhin and A.~Varchenko.
\newblock In {\em Geometry, Topology, and Physics}, page 293 (1995).

\bibitem{Amico01a}
L.~Amico, A.~Di Lorenzo, and A.~Osterloh, 
Phys.~Rev.~Lett. {\bf 86}, 5759 (2001).

\bibitem{Amico01b}
L.~Amico, A.~Di Lorenzo, and A.~Osterloh, 
Nucl.~Phys.~B {\bf 614}, 449  (2001).

\bibitem{Dukelsky01}
J.~Dukelsky, C.~Esebbag, and P.~Schuck, 
Phys.~Rev.~Lett. {\bf 87}, 66403 (2001).

\bibitem{vonDelft01}
J.~von Delft and R.~Poghossian, 
{\em Algebraic Bethe Ansatz for a  discrete-state BCS pairing model}, cond-mat/0106405.

\bibitem{Sklyanin88}
E.~K. Sklyanin, 
  J.~Phys.~A {\bf 21}, 2375 (1988).

\bibitem{Hikami95}
K. Hikami, 
J.~Phys.~A {\bf 28}, 4997 (1995).

\bibitem{Hikami95a}
K.~Hikami, J.~Phys.~A {\bf 28}, 4053 (1995).

\bibitem{Cherednik84}
I.~Cherednik, Theor. Math. Phys. {\bf 61}, 35 (1984).

\bibitem{deVega84}
H.~J. de~Vega, 
 Nucl.~Phys.~B {\bf 240}, 495 (1984).

\bibitem{Babujian82}
H.~M. Babujian, 
Phys.~Lett.~A {\bf 90}, 479 (1982).

\bibitem{Takhtajan82}
L.~A. Takhtajan, 
Phys.~Lett.~A {\bf 87}, 479 (1982).

\bibitem{Electro2002}
L.~Amico, A.~Di Lorenzo, A.~Mastellone, A.~Osterloh, and R.~Raimondi, 
Annals of Physics {\bf 299}, 228 (2002).

\bibitem{Dilorenzo01}
A. Di Lorenzo,
\newblock {\em A new class of exactly solvable models}.
\newblock PhD thesis, Universit\`a di Catania, Italy, (2001).

\bibitem{Richardson-private}
R.W. Richardson.
\newblock {\em private communication}.

\bibitem{Fateev80}
V.I. Fateev and A.B. Zamolodchikov, Sov. J. Nucl. Phys. {\bf 32}, 298 (1980).

\bibitem{Sogo83}
K.~Sogo, Y.~Akutsu, and T.~Abe, Prog. Theor. Phys. {\bf 70}, 730 (1983).

\bibitem{Kulish81}
P.P. Kulish, N.~Reshetikhin, and E.K. Sklyanin, Lett. Math. Phys. {\bf 5}, 393
  (1981).

\bibitem{Kulish01}
P. P. Kulish and N. Manojlovic,
Lett. Math. Phys. {\bf 55}, 77 (2001).

\bibitem{DeVega93}
H.~J. de~Vega and A.~Gonzalez-Ruiz, J. Phys. A {\bf 26}, L519 (1993).

\bibitem{Ghoshal94}
S. Ghoshal and A. Zamolodchikov, Int. J. Mod. Phys. A {\bf 9}, 3841
(1994).


\bibitem{DeVega94}
H. J. de Vega and A. Gonz\'alez-Ruiz,
Mod. Phys. Lett. A {\bf 9}, 2207 (1994).

\bibitem{Doikou98}
A. Doikou and R. I. Nepomechie,
Nucl. Phys. B {\bf 530}, 641 (1998).
\end{thebibliography}
\end{document}